\documentclass[twocolumn,showpacs,showkeys]{revtex4}
\usepackage{graphicx}
\usepackage{bm}
\usepackage{color}
\usepackage{amsmath}
\usepackage{natbib}

\begin{document}

\title{Oblique propagation of longitudinal waves in magnetized spin-1/2 plasmas: Independent evolution of spin-up and spin-down electrons}

\author{Pavel A. Andreev}
\email{andreevpa@physics.msu.ru}
\author{L. S. Kuz'menkov}%
\email{lsk@phys.msu.ru}
\affiliation{Faculty of physics, Lomonosov Moscow State University, Moscow, Russian Federation.}

 \date{\today}

\begin{abstract}
We consider quantum plasmas of electrons and motionless ions. We describe separate evolution of spin-up and spin-down electrons. We present corresponding set of quantum hydrodynamic equations. We assume that plasmas are placed in an uniform external magnetic field. We account different occupation of spin-up and spin-down quantum states in equilibrium degenerate plasmas. This effect is included via equations of state for pressure of each species of electrons. We study oblique propagation of longitudinal waves. We show that instead of two well-known waves (the Langmuir wave and the Trivelpiece--Gould wave), plasmas reveal four wave solutions. New solutions exist due to both the separate consideration of spin-up and spin-down electrons \textit{and} different occupation of spin-up and spin-down quantum states in equilibrium state of degenerate plasmas.
\end{abstract}


\maketitle




\section{\label{sec:level1} Introduction}

Spin evolution in quantum plasmas \cite{Maksimov VestnMSU 2000}, \cite{Maksimov Izv 2000}, \cite{MaksimovTMP 2001}, \cite{Marklund PRL07}, \cite{Brodin NJP 07}, \cite{Andreev RPJ 07} leads to existence of new waves \cite{Vagin Izv RAN 06}, \cite{Andreev VestnMSU 2007}, \cite{Andreev AtPhys 08}, \cite{Misra JPP 10}, \cite{Andreev IJMP 12}. Difference of population of spin-up and spin-down quantum states in equilibrium plasmas, which can be accounted by corresponding equation of state, also gives contribution in dispersion of plasma waves \cite{Andreev spin-up and spin-down 1405}, \cite{Andreev 1403 exchange}, \cite{Trukhanova 1405 exchange}. Moreover, consideration of separate evolution of spin-up and spin-down electrons reveal existence of new longitudinal wave \cite{Andreev spin-up and spin-down 1405}. The new longitudinal wave was obtained at consideration of wave propagation parallel and perpendicular to an external magnetic field. Its existence is related to different population of spin-up and spin-down quantum states in equilibrium plasmas.

If we have equal population of spin-up and spin-down quantum states and consider ions as motionless background we have one longitudinal wave. It is the Langmuir wave. Presence of an external magnetic field reveals in anisotropy of the Langmuir wave dispersion. Moreover, if we consider propagation of longitudinal waves parallel or perpendicular to the external magnetic field we have the Langmuir wave only. However, the second wave exists at oblique propagation. It is well-known Trivelpiece--Gould wave. Appearance of the second wave solution at oblique propagation of the longitudinal waves in plasmas encourages us to check existence of new oblique propagating waves in plasmas at different population of spin-up and spin-down quantum states.

In this paper we present further application of separated spin evolution QHD (SSE-QHD). We consider spin-up electrons and spin-down electrons as two different species. Corresponding QHD equations were directly derived from the Pauli equation \cite{Andreev spin-up and spin-down 1405}. Let us mention a couple of papers \cite{Iqbal JPP 13}, \cite{Shahid PP 12}, where attempts to suggest separated spin evolution QHD were made.

Some results on spinless quantum plasmas and quantum plasmas of spin-1/2 particles were reviewed in Refs. \cite{Shukla PhUsp 2010}-\cite{Uzdensky arxiv review 14}. We would like to mention several results obtained in the field of spin-1/2 quantum plasmas in past years. Contribution of the Coulomb exchange and spin-spin exchange interactions in spectrums of magnetized spin-1/2 quantum plasmas was described in Ref. \cite{Andreev AtPhys 08}. Propagation of neutron beam via magnetized spin-1/2 quantum plasmas and generation of waves by neutron beams were also considered in Ref. \cite{Andreev AtPhys 08}. Explicit account of the spin-current interaction by means of the many-particle quantum hydrodynamic (MPQHD) method in spin-1/2 quantum plasmas was performed in Ref. \cite{Andreev RPJ 07}. The spin-current interaction is the interaction between magnetic moments related to spins and electric current of moving charges in plasmas occurring by means the magnetic field. The spin-orbit interaction and its influence on spectrums of plasma waves, spin-plasma waves and processes of neutron beam--magnetized spin-1/2 quantum plasmas interaction \cite{Andreev IJMP 12}. Consistent consideration of the quantum Bohm potential in system of spinning particles in terms of the MPQHD was done in Ref. \cite{Andreev Asenjo 13}. Development of general problems in modeling of collective behavior of spinning particle quantum plasmas has been performed. For instance, the gauge-free Hamiltonian structure of an extended kinetic theory, for which the intrinsic spin of the particles is taken into account was developed in Ref. \cite{Marklund PLA 11}. Model the neutron fluid as a spin quantum plasma where the electromagnetic interaction is trough the magnetic moment of the neutron is presented in Ref. \cite{Mahajan PL A 13}, as an excellent application of QHD to systems of neutral particles with spin. An extended vorticity evolution equation for the quantum spinning plasma was considered in Refs. \cite{Mahajan PL A 13} and \cite{Trukhanova PrETP 13}. The influence of the intrinsic spin of electrons on whistler mode was also investigated in Ref. \cite{Trukhanova PrETP 13}. Effects of the spin and the Bohm potential in the oblique propagation of magnetosonic waves were considered in Ref. \cite{Asenjo PL A 12}.
From QHD description with intrinsic magnetization, a new plasma instability was obtained in Ref. \cite{Bychkov PP 10}. It was shown that the instability develops in a nonuniform plasma when the electron concentration and temperature vary along an externally applied magnetic field. Authors obtained that Alfven waves play an important role in the instability. Linear and nonlinear relations for slow and fast magnetosonic modes were derived in Refs. \cite{Mushtaq PP 10} and \cite{Mushtaq PP EPJD 11}, where spin effects are incorporated via spin force and macroscopic spin magnetization current. Their solution shows a general shock wave profile superposed by a perturbative solitary-wave contribution \cite{Mushtaq PP 10}. Magnetosonic waves were studied in magnetized degenerate electron-positron-ion plasmas with spin effects \cite{Mushtaq PP 12}. It was demonstrated that the effect of quantum corrections in the presence of positron concentration significantly modifies the dispersive properties of these modes.
The magnetosonic waves and their interactions in spin-1/2 degenerate quantum plasmas were investigated in Ref. \cite{Chang Li PP 14}. Electron spin -1/2 effects on the parametric decay instability of oblique Langmuir wave into low-frequency electromagnetic shear Alfven wave and left-handed circularly polarized wave was considered in Ref. \cite{Shahid PP 13}. The effect of spin induced magnetization on Jeans instability of quantum plasmas was studied in Ref. \cite{Sharma PP 14}. Effects of electron spin on the kinetic Alfven waves in the presence of uniform static magnetic field were studied in Ref. \cite{Hussain PLA 13}. It was demonstrated  that the kinetic Alfven wave frequency decreases due to the electron spin contribution in the kinetic limit while in the inertial limit they are almost unaffected in a hot magnetized plasma.

This paper is organized as follows.  In Sec. II model of separated spin evolution QHD is presented and described. In Sec. III dispersion of longitudinal waves in quantum plasmas with different population of spin-up and spin-down quantum states. We show existence of four wave solutions instead of two well-known solutions. In Sec. IV brief summary of obtained results is presented.

\section{\label{sec:level1} Model}

We should start derivation of the SSE-QHD equations from many-particle Pauli equation with explicit account of interparticle interactions \cite{MaksimovTMP 2001}, \cite{Andreev RPJ 07}, \cite{Andreev Asenjo 13}, \cite{MaksimovTMP 1999}, \cite{Andreev PRB 11}. However essential part can be found from the single particle Pauli equation \cite{Brodin NJP 07}, \cite{Takabayasi PTP 55 a}, \cite{Takabayasi PTP 54}, \cite{Takabayasi PTP 55 b}.

The Pauli equation
\begin{equation}\label{SUSD Obl Pauli} \imath\hbar\partial_{t}\psi=\biggl(\frac{(\widehat{\textbf{p}}-\frac{q_{e}}{c}\textbf{A})^{2}}{2m}+q_{e}\varphi-\gamma_{e}\widehat{\mbox{\boldmath $\sigma$}} \textbf{B}\biggr)\psi \end{equation}
governs evolution of spinor wave function $\psi(\textbf{r},t)$, where $\varphi=\varphi_{ext}$, $\textbf{A}=\textbf{A}_{ext}$ are the scalar and vector potentials of external electromagnetic fields, $\textbf{B}=\textbf{B}_{ext}$ is the external magnetic field, $q_{e}=-e$ is the charge of electron, $m$ is the mass of the particle under consideration, $\gamma_{e}$ is the gyromagnetic ratio, $\widehat{\textbf{p}}=-\imath\hbar\nabla$ is the momentum operator, $\nabla$ is the gradient operator, $\mbox{\boldmath $\sigma$}$ is the vector of Pauli matrixes, $\hbar$ is the reduced Planck constant, $c$ is the speed of light, $\widehat{\mbox{\boldmath $\sigma$}}$ is the vector constructed of
the Pauli matrixes
\begin{equation}\label{SUSD Obl}\begin{array}{ccc} \widehat{\sigma}_{x}=\left(\begin{array}{ccc}0& 1\\
1& 0\\
\end{array}\right),&
\widehat{\sigma}_{y}=\left(\begin{array}{ccc}0& -\imath \\
\imath & 0 \\
\end{array}\right),&
\widehat{\sigma}_{z}=\left(\begin{array}{ccc} 1& 0\\
0& -1\\
\end{array}\right).
\end{array}\end{equation}

The spinor wave function $\psi$ can be presented as
\begin{equation}\label{SUSD Obl spinor wave function}
\psi=
\left(\begin{array}{ccc}
\psi_{\uparrow} \\
\psi_{\downarrow} \\
\end{array}\right).\end{equation}
Applying wave functions describing spin-up $\psi_{\uparrow}$ and spin-down $\psi_{\downarrow}$ states we can write probability density to find the particle in a point $\textbf{r}$ with spin-up $\rho_{\uparrow}=\mid\psi_{\uparrow}\mid^{2}$ or spin-down $\rho_{\downarrow}=\mid\psi_{\downarrow}\mid^{2}$. We also see $\rho=\rho_{\uparrow}+\rho_{\downarrow}$. Directions up $\uparrow$ (down $\downarrow$) corresponds to spins having same (opposite) direction as (to) the external magnetic field. While magnetic moments have opposite to spin directions.

The spin density $S_{z}$ of electrons is the difference between concentrations of electrons with different projection of spin $S_{z}=\psi^{+}\sigma_{z}\psi =\rho_{\uparrow}-\rho_{\downarrow}$. We have that the z-projection of the spin density $S_{z}$ is not an independent variable in this representation of the quantum hydrodynamics.

In many-particle systems we have concentration of particles $n(\textbf{r},t)$, which are proportional to the probability density to find each particle in the point $\textbf{r}$, hence we have $n_{\uparrow}=\langle\rho_{\uparrow}\rangle$, $n_{\downarrow}=\langle\rho_{\downarrow}\rangle$, and $n=n_{\uparrow}+n_{\downarrow}$.

Applying the explicit form of the Pauli matrixes we can rewrite the Pauli equation (\ref{SUSD Obl Pauli}) in more explicit form, in terms of $\psi_{\uparrow}$ and $\psi_{\downarrow}$ (see equations (4) and (5) in Ref. \cite{Andreev spin-up and spin-down 1405}). These equations allow to derive equations for $n_{\uparrow}$, $\textbf{v}_{\uparrow}$ \emph{and} $n_{\downarrow}$, $\textbf{v}_{\downarrow}$. These equations were obtained in Ref. \cite{Andreev spin-up and spin-down 1405}. Here we present and apply these equations.

\begin{figure}
\includegraphics[width=8cm,angle=0]{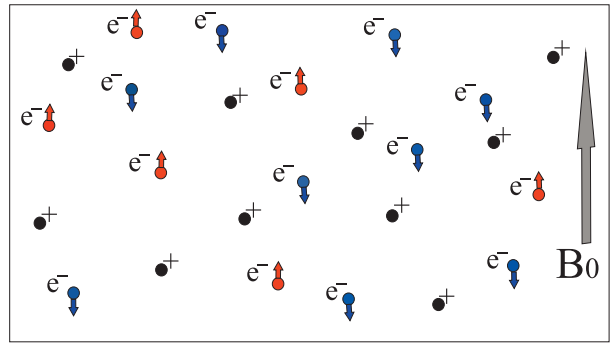}
\caption{\label{SUSD Obl F0a} (Color online) The figure shows system of spin-up and spin-down electrons on neutralizing ion background. Spin-up (spin-down) electrons are presented by red (blue) circles with arrows directed in the same directions as (in opposite direction to) the external magnetic field. The external magnetic field is presented by the large grey arrow. Black circles with symbol $+$ present ions.}
\end{figure}

The continuity equations appear for each species of electrons
\begin{equation}\label{SUSD Obl cont eq electrons spin UP}
\partial_{t}n_{\uparrow}+\nabla(n_{\uparrow}\textbf{v}_{\uparrow})=\frac{\gamma}{\hbar}(B_{y}S_{x}-B_{x}S_{y}) \end{equation}
for spin0up electrons, and
\begin{equation}\label{SUSD Obl cont eq electrons spin DOWN}
\partial_{t}n_{\downarrow}+\nabla(n_{\downarrow}\textbf{v}_{\downarrow})=\frac{\gamma}{\hbar}(B_{x}S_{y}-B_{y}S_{x}) \end{equation}
for spin-down electrons.

In the continuity equations we have the following physical quantities: $n_{\uparrow}$ ($n_{\downarrow}$) is the concentration of electrons baring spin-up (spin-down), $\textbf{v}_{\uparrow}$ ($\textbf{v}_{\downarrow}$) is the velocity field of electrons baring spin-up (spin-down), $S_{x}$ and $S_{y}$ are projections of the spin density vector.

The right-hand side of the continuity equations exist due to the spin-spin interaction between electrons. Numbers of spin-up and spin-down electrons do not conserve due to spin-spin interaction. Total number of electrons conserves only.

We also have the couple of vector Euler equations. These equations describe evolution of the momentum density in each species of electrons.
$$mn_{\uparrow}(\partial_{t}+\textbf{v}_{\uparrow}\nabla)\textbf{v}_{\uparrow}+\nabla p_{\uparrow}-\frac{\hbar^{2}}{4m}n_{\uparrow}\nabla\Biggl(\frac{\triangle n_{\uparrow}}{n_{\uparrow}}-\frac{(\nabla n_{\uparrow})^{2}}{2n_{\uparrow}^{2}}\Biggr)$$
$$=q_{e}n_{\uparrow}\biggl(\textbf{E}+\frac{1}{c}[\textbf{v}_{\uparrow},\textbf{B}]\biggr)+\frac{\gamma_{e}}{m}n_{\uparrow}\nabla B_{z}$$
\begin{equation}\label{SUSD Obl Euler eq electrons spin UP} +\frac{\gamma_{e}}{2m}(S_{x}\nabla B_{x}+S_{y}\nabla B_{y})+\frac{\gamma_{e}}{\hbar}(\textbf{J}_{(M)x}B_{y}-\textbf{J}_{(M)y}B_{x}),\end{equation}
and
$$mn_{\downarrow}(\partial_{t}+\textbf{v}_{\downarrow}\nabla)\textbf{v}_{\downarrow}+\nabla p_{\downarrow}-\frac{\hbar^{2}}{4m}n_{\downarrow}\nabla\Biggl(\frac{\triangle n_{\downarrow}}{n_{\downarrow}}-\frac{(\nabla n_{\downarrow})^{2}}{2n_{\downarrow}^{2}}\Biggr)$$
$$=q_{e}n_{\downarrow}\biggl(\textbf{E}+\frac{1}{c}[\textbf{v}_{\downarrow},\textbf{B}]\biggr)-\frac{\gamma_{e}}{m}n_{\downarrow}\nabla B_{z}$$
\begin{equation}\label{SUSD Obl Euler eq electrons spin DOWN}+\frac{\gamma_{e}}{2m}(S_{x}\nabla B_{x}+S_{y}\nabla B_{y})+\frac{\gamma_{e}}{\hbar}(\textbf{J}_{(M)y}B_{x}-\textbf{J}_{(M)x}B_{y}),\end{equation}
with
\begin{equation}\label{SUSD Obl Spin current x} \textbf{J}_{(M)x}=\frac{1}{2}(\textbf{v}_{\uparrow}+\textbf{v}_{\downarrow})S_{x}-\frac{\hbar}{4m} \biggl(\frac{\nabla n_{\uparrow}}{n_{\uparrow}}+\frac{\nabla n_{\downarrow}}{n_{\downarrow}}\biggr)S_{y}, \end{equation}
and
\begin{equation}\label{SUSD Obl Spin current y} \textbf{J}_{(M)y}= \frac{1}{2}(\textbf{v}_{\uparrow}+\textbf{v}_{\downarrow})S_{y}+\frac{\hbar}{4m}\biggl(\frac{\nabla n_{\uparrow}}{n_{\uparrow}}+\frac{\nabla n_{\downarrow}}{n_{\downarrow}}\biggr)S_{x}, \end{equation}
where $q_{e}=-e$, $\gamma_{e}=-g\frac{e\hbar}{2mc}$ is the gyromagnetic ratio for electrons, and $g=1+\alpha/(2\pi)=1.00116$, where $\alpha=1/137$ is the fine structure constant, gets into account the anomalous magnetic moment of electron. $\textbf{J}_{(M)x}$ and $\textbf{J}_{(M)y}$ are elements of the spin current tensor $J^{\alpha\beta}$.

The last group of terms in the Euler equations exist due to nonconservation of numbers of spin-up and spin-down electrons. Hence these terms are related to the spin-spin interaction.

\begin{figure}
\includegraphics[width=8cm,angle=0]{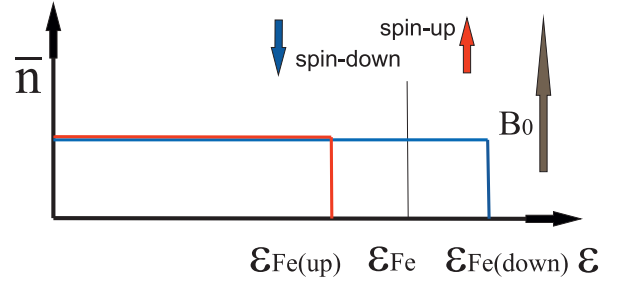}
\caption{\label{SUSD Obl F0b} (Color online) The figure shows distribution functions $\overline{n}$ of degenerate spin-up and spin-down electrons being in external magnetic field. This distribution function gives average occupation number of quantum states with different energies.}
\end{figure}

Equations (\ref{SUSD Obl cont eq electrons spin UP}) -(\ref{SUSD Obl Spin current y}) contain $S_{x}$ and $S_{y}$. Equations of evolution of $S_{x}$ and $S_{y}$ were derived in Ref. \cite{Andreev spin-up and spin-down 1405}. We also present them here to have closed set of the SSE-QHD. $S_{x}=\psi^{*}\sigma_{x}\psi=\psi_{\downarrow}^{*}\psi_{\uparrow}+\psi_{\uparrow}^{*}\psi_{\downarrow}$, $S_{y}=\psi^{*}\sigma_{y}\psi=\imath(\psi_{\downarrow}^{*}\psi_{\uparrow}-\psi_{\uparrow}^{*}\psi_{\downarrow})$. $S_{x}$ and $S_{y}$ appear as mixed combinations of $\psi_{\uparrow}$ and $\psi_{\downarrow}$. These quantities do not related to different species of electrons having different spin direction. $S_{x}$ and $S_{y}$ describe simultaneous evolution of both species.

Equations of transverse spin projection evolution $S_{x}$ and $S_{y}$ appear as follows
$$\partial_{t}S_{x}+\frac{1}{2}\nabla[S_{x}(\textbf{v}_{\uparrow}+\textbf{v}_{\downarrow})]$$
\begin{equation}\label{SUSD Obl eq for Sx} -\frac{\hbar}{4m}\nabla\Biggl(S_{y}\biggl(\frac{\nabla n_{\uparrow}}{n_{\uparrow}}-\frac{\nabla n_{\downarrow}}{n_{\downarrow}}\biggr)\Biggr)=\frac{2\gamma_{e}}{\hbar}\biggl(B_{z}S_{y}-B_{y}(n_{\uparrow}-n_{\downarrow})\biggr),\end{equation}
and
$$\partial_{t}S_{y}+\frac{1}{2}\nabla[S_{y}(\textbf{v}_{\uparrow}+\textbf{v}_{\downarrow})]$$
\begin{equation}\label{SUSD Obl eq for Sy} +\frac{\hbar}{4m}\nabla\Biggl(S_{x}\biggl(\frac{\nabla n_{\uparrow}}{n_{\uparrow}}-\frac{\nabla n_{\downarrow}}{n_{\downarrow}}\biggr)\Biggr)=\frac{2\gamma_{e}}{\hbar}\biggl(B_{x}(n_{\uparrow}-n_{\downarrow})-B_{z}S_{x}\biggr).\end{equation}

In this paper we are focused on the longitudinal waves. Hence we present quasi-electrostatic set of the Maxwell equations
\begin{equation}\label{SUSD Obl div E} \nabla \textbf{E}=4\pi\biggl(en_{i}-en_{e\uparrow}-en_{e\downarrow}\biggr),\end{equation}
and
\begin{equation}\label{SUSD Obl ror E} \nabla\times \textbf{E}=0.\end{equation}

To get closed set of QHD equations we apply the following equation of state for each species of electrons
\begin{equation}\label{SUSD Obl EqState partial}p_{s}=\frac{(6\pi^{2})^{2/3}}{5}\frac{\hbar^{2}}{m}n_{s}^{5/3},\end{equation}
where $s=\uparrow$ or $\downarrow$.

We show below that difference between $p_{\uparrow}$ and $p_{\downarrow}$ due to difference of $n_{\uparrow}$ and $n_{\downarrow}$ leads to new effects in quantum plasmas.

\begin{figure}
\includegraphics[width=8cm,angle=0]{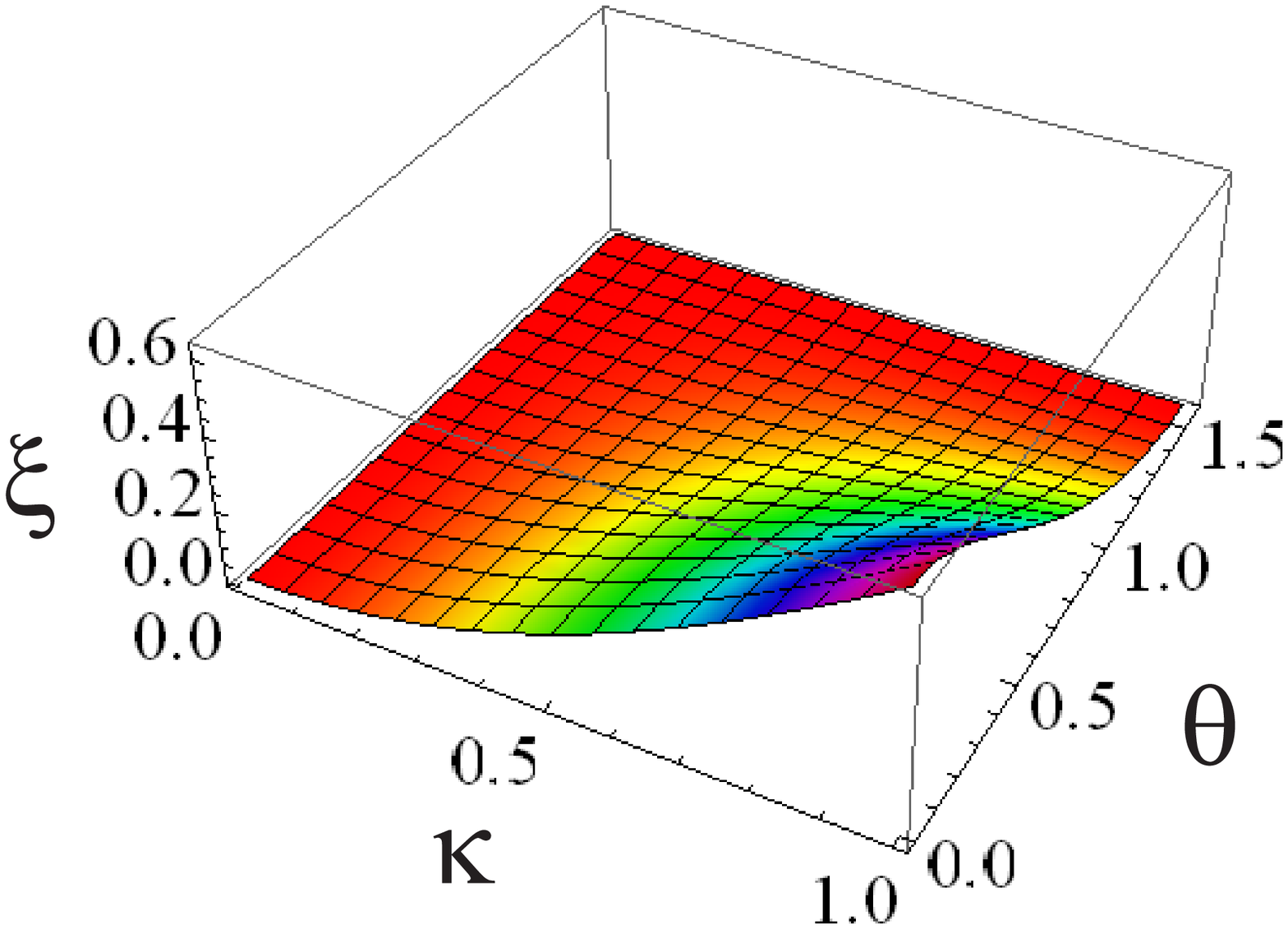}
\caption{\label{SUSD Obl F1} (Color online) This figure shows dispersion surface of the longitudinal wave existing at $\theta=0$.}
\end{figure}

\begin{figure}
\includegraphics[width=8cm,angle=0]{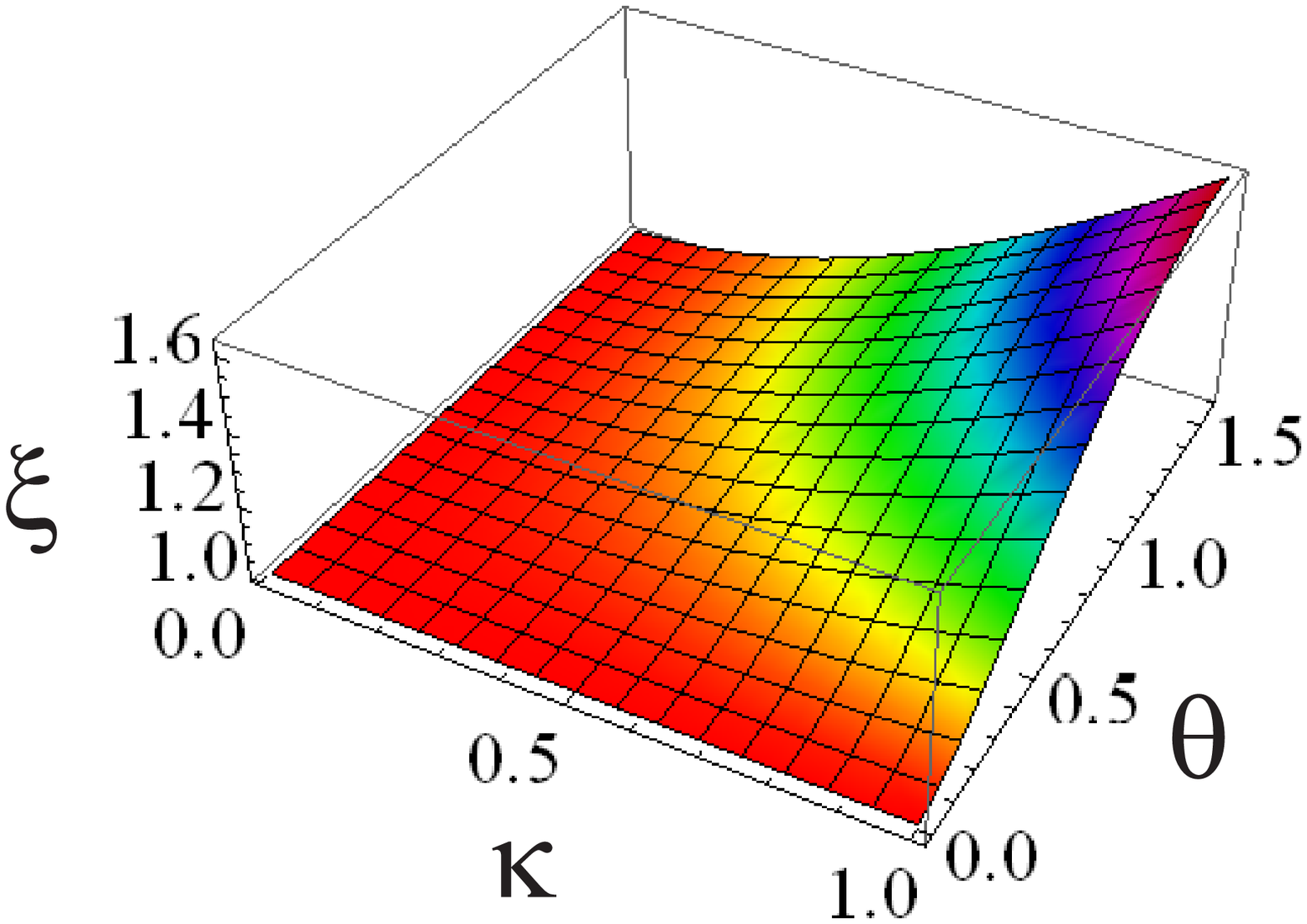}
\caption{\label{SUSD Obl F3} (Color online) This figure shows dispersion surface of the longitudinal wave existing at $\theta=\pi/2$.}
\end{figure}

A longitudinal wave propagating parallel to the external magnetic field was discovered in Ref. \cite{Andreev spin-up and spin-down 1405}. A longitudinal wave propagating perpendicular to the external magnetic field was also obtained in Ref. \cite{Andreev spin-up and spin-down 1405}. Their dispersion dependencies differ from each other by a constant $\omega_{\perp}^{2}(k)=\omega_{\parallel}^{2}(k)+\Omega^{2}$. In paper \cite{Andreev spin-up and spin-down 1405} these solutions were interpreted as part of one dispersion surface $\omega(k,\theta)$ existing at $\theta=0$ and $\theta=\pi/2$. From figures (\ref{SUSD Obl F1}) and (\ref{SUSD Obl F3}) we see that these two solutions are related to different dispersion surfaces. Figure (\ref{SUSD Obl F1}) (figure (\ref{SUSD Obl F3})) shows dispersion surface of wave existing at $\theta=0$ ($\theta=\pi/2$).

\section{Dispersion of longitudinal waves}

\begin{figure}
\includegraphics[width=8cm,angle=0]{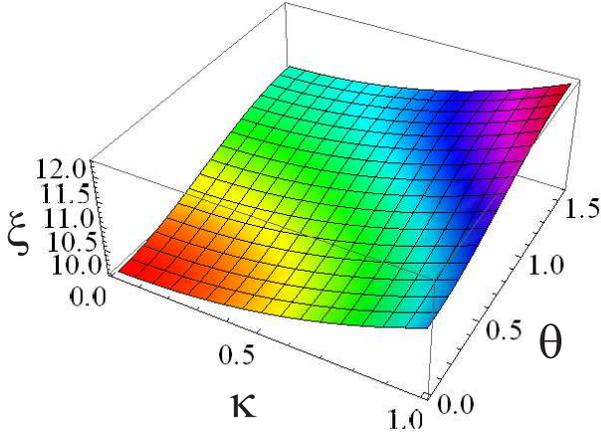}
\caption{\label{SUSD Obl F4} (Color online) The figure shows anisotropic dispersion dependence of the Langmuir wave.}
\end{figure}

\begin{figure}
\includegraphics[width=8cm,angle=0]{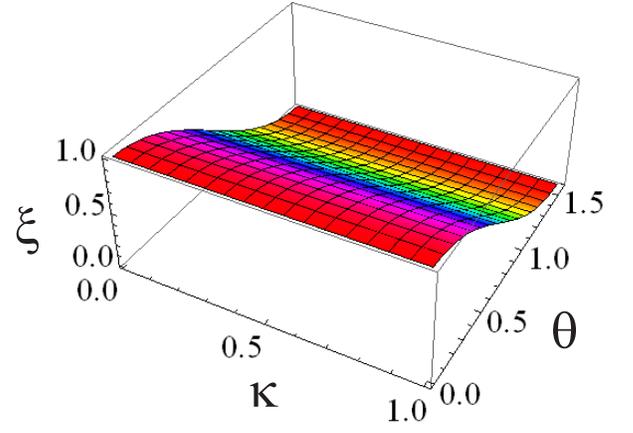}
\caption{\label{SUSD Obl F2} (Color online) Trivelpiece--Gould wave dispersion is presented on the figure.}
\end{figure}

Equilibrium condition is described by the non-zero concentrations $n_{0\uparrow}$, $n_{0\downarrow}$, $n_{0}=n_{0\uparrow}+n_{0\downarrow}$, and external magnetic field $\textbf{B}_{ext}=B_{0}\textbf{e}_{z}$. Other quantities equal to zero $\textbf{v}_{0\uparrow}=\textbf{v}_{0\downarrow}=0$, $\textbf{E}_{0}=0$, $S_{0x}=S_{0y}=0$.

Difference of spin-up and spin-down concentrations of electrons $\Delta n=n_{0\uparrow}-n_{0\downarrow}$ is caused  by external magnetic field. Since electrons are negative their spins get preferable direction opposite to the external magnetic field $\frac{\Delta n}{n_{0}}=\tanh\biggl(\frac{\gamma_{e}B_{0}}{\varepsilon_{Fe}}\biggr)=-\tanh\biggl(\frac{\mid\gamma_{e}\mid B_{0}}{\varepsilon_{Fe}}\biggr)$, where $\varepsilon_{Fe}=(3\pi^{2})^{\frac{2}{3}}\frac{\hbar^{2}}{2m}n^{\frac{2}{3}}_{0}$ is the Fermi energy.

Assuming that perturbations $\delta n_{\uparrow}$,
$\delta n_{\downarrow}$,
$\delta \textbf{v}_{\uparrow}$,
$\delta \textbf{v}_{\downarrow}$,
$\delta \textbf{E}$,
$\delta \textbf{B}$,
$\delta S_{x}$,
$\delta S_{y}$ are monochromatic
\begin{equation}\label{SUSD Obl perturbations}
\delta f=F_{A}e^{-\imath\omega t+\imath \textbf{k} \textbf{r}},\end{equation}
where $\delta f$ presents perturbations of physical quantities, and $F$ is corresponding amplitude.
we get a set of linear algebraic equations relatively to amplitudes of perturbations. Condition of existence of nonzero solutions for amplitudes of perturbations gives us a dispersion equation.

We assume that $\textbf{k}=\{k_{x}, 0, k_{z}\}$ and $k_{x}=k\sin\theta$, $k_{z}=k\cos\theta$, where $k=\sqrt{k_{x}^{2}+k_{z}^{2}}$, and $\theta$ is the angle between direction of wave propagation and direction of the external magnetic field.

For longitudinal waves we have that perturbations of magnetic field equal to zero $\delta \textbf{B}=0$.

\begin{equation}\label{SUSD Obl definition of U} U_{s}^{2}=\frac{(6\pi^{2})^{\frac{2}{3}}}{3}\frac{\hbar^{2}}{m^{2}}n_{0s}^{\frac{2}{3}}=\frac{2^{\frac{2}{3}}}{3}v_{Fe(s)}^{2},\end{equation}
with $s=\uparrow$ or $\downarrow$.

Equations (\ref{SUSD Obl eq for Sx}) and (\ref{SUSD Obl eq for Sy}) describe precession of spins around the external magnetic field. Frequency of precession is $\omega_{pr}=\frac{2\mid\gamma_{e}\mid}{\hbar}B_{0}$ It does not affect matter waves described by the continuity and Euler equations.

The longitudinal waves are described by the continuity (\ref{SUSD Obl cont eq electrons spin UP}), (\ref{SUSD Obl cont eq electrons spin DOWN}) and Euler (\ref{SUSD Obl Euler eq electrons spin UP}), (\ref{SUSD Obl Euler eq electrons spin DOWN}) equation of material fields \textit{and} equations of the electric field (\ref{SUSD Obl div E}), (\ref{SUSD Obl ror E}). These equations lead to the following dispersion equation
$$1-\biggl(\frac{\sin^{2}\theta}{\omega^{2}-\Omega^{2}}+\frac{\cos^{2}\theta}{\omega^{2}}\biggr)\times$$
$$\times\Biggl[\frac{\omega_{Le\uparrow}^{2}}{1-(\frac{\sin^{2}\theta}{\omega^{2}-\Omega^{2}}+\frac{\cos^{2}\theta}{\omega^{2}})(U_{\uparrow}^{2}+\frac{\hbar^{2}k^{2}}{4m^{2}})k^{2}}$$
\begin{equation}\label{SUSD Obl Longit disp eq general} +\frac{\omega_{Le\downarrow}^{2}}{1-(\frac{\sin^{2}\theta}{\omega^{2}-\Omega^{2}}+\frac{\cos^{2}\theta}{\omega^{2}})(U_{\downarrow}^{2}+\frac{\hbar^{2}k^{2}}{4m^{2}})k^{2}}\Biggr]=0\end{equation}
where $\omega_{Le\uparrow}^{2}=\frac{4\pi e^{2}n_{0\uparrow}}{m}$, and $\omega_{Le\downarrow}^{2}=\frac{4\pi e^{2}n_{0\downarrow}}{m}$ are the Langmuir frequencies for spin-up and spin-down electrons. $\omega_{Le\uparrow}^{2}$ and $\omega_{Le\downarrow}^{2}$ are partial Langmuir frequencies. Their sum $\omega_{Le}^{2}=\omega_{Le\uparrow}^{2}+\omega_{Le\downarrow}^{2}$ gives full Langmuir frequency of the system.

Dispersion equation (\ref{SUSD Obl Longit disp eq general}) is an equation of fourth degree on the frequency square $\omega^{2}$. Hence we can expect existence of four waves, whereas there is two well-known longitudinal waves in magnetized three dimensional electron gas. They are the Langmuir and the Trivelpiece--Gould wave.

At $\theta=0$ equation (\ref{SUSD Obl Longit disp eq general}) simplifies to
\begin{equation}\label{SUSD Obl Longit disp eq} 1-\frac{\omega_{Le\uparrow}^{2}}{\omega^{2}-(U_{\uparrow}^{2}+\frac{\hbar^{2}k^{2}}{4m^{2}})k^{2}}
-\frac{\omega_{Le\downarrow}^{2}}{\omega^{2}-(U_{\downarrow}^{2}+\frac{\hbar^{2}k^{2}}{4m^{2}})k^{2}}=0.\end{equation}

If occupations of states equal to each other than we have $n_{0\uparrow}=n_{0\downarrow}$. As consequence we get $U_{\uparrow}^{2}=U_{\downarrow}^{2}=\frac{1}{3}v_{Fe}^{2}$, $v_{Fe}=(3\pi^{2}n_{0})^{\frac{1}{3}}\hbar/m$, and $\omega_{Le\uparrow}^{2}=\omega_{Le\downarrow}^{2}=\frac{1}{2}\omega_{Le}^{2}$. In this limit new solutions do not appear.

Equation (\ref{SUSD Obl Longit disp eq}) has two solutions. One of them is the Langmuir wave. The second branch was discovered in Ref. \cite{Andreev spin-up and spin-down 1405}. Analytical analysis of spectrum of new wave was presented in Ref. \cite{Andreev spin-up and spin-down 1405}. Difference in occupation of spin-up and spin-down quantum states by electrons gives a contribution in the dispersion of Langmuir waves. Analytical expressions for this contribution is obtained in Ref. \cite{Andreev spin-up and spin-down 1405}. In this paper we present numerical analysis of this effect at oblique propagation of the Langmuir wave.

\begin{figure}
\includegraphics[width=8cm,angle=0]{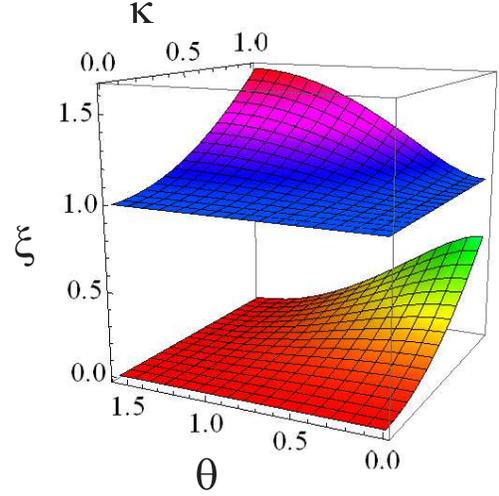}
\caption{\label{SUSD Obl F5} (Color online) Comparison of two new branches of wave dispersion is depicted on the figure.}
\end{figure}

\begin{figure}
\includegraphics[width=8cm,angle=0]{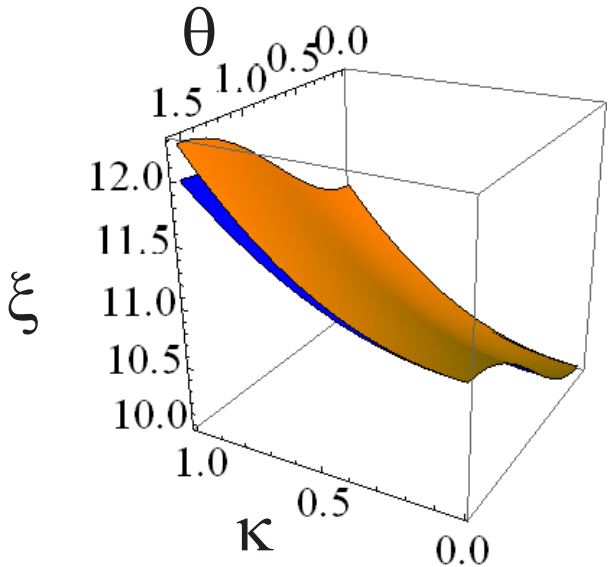}
\caption{\label{SUSD Obl F6} (Color online) Comparison of the Langmuir wave dispersion surfaces in two cases: 1)when we consider separated spin evolution and different Fermi pressure for two species of electrons (upper surface); 2) when we do not account these effects (lower surface) is presented.}
\end{figure}

\begin{figure}
\includegraphics[width=8cm,angle=0]{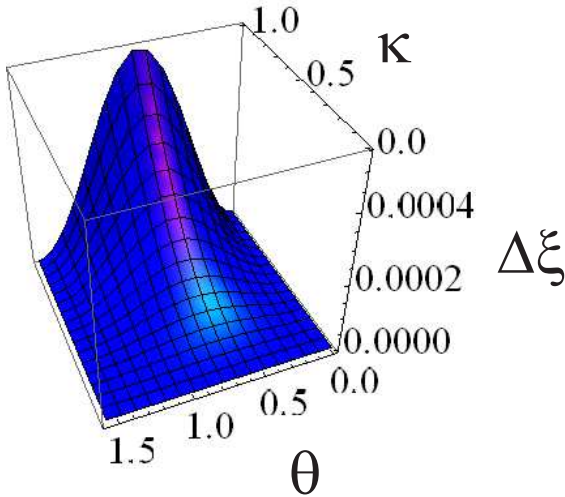}
\caption{\label{SUSD Obl F7} (Color online) Shift of the Trivelpiece--Gould wave dispersion at account of separated spin evolution and different Fermi pressure for two species of electrons is presented.}
\end{figure}

\subsection{Numerical analysis}

In this subsection we present numerical analysis of dispersion equation (\ref{SUSD Obl Longit disp eq general}).
We assume that the equilibrium particle concentration has the following value
$n_{0}=10^{22}$ cm$^{-3}$. We also assume that the Langmuir frequency and the cyclotron frequency are related as $\omega_{Le}^{2}=10\Omega^{2}$. Consequently we have the following magnitude of the external magnetic field $B_{0}=3\sqrt{4\pi}10^{7}$G.

During numerical analysis of the dispersion equation we apply dimensionless wave vector module
$\kappa=\frac{kU_{av}}{\mid \Omega\mid}=\frac{(3\pi^{2})^{\frac{1}{3}}}{\sqrt{3}}\frac{\hbar c}{eB_{0}}n_{0}^{\frac{1}{3}}k$
and dimensionless frequency square
$\xi=\frac{\omega^{2}}{\Omega^{2}}$.

This particle concentration corresponds to electrons in metals. The quantum Bohm potential is essential at much higher densities existing in astrophysical objects. Hence, at numerical analysis  we do not include contribution of the quantum Bohm potential.

Figs. (\ref{SUSD Obl F4})-(\ref{SUSD Obl F3}) show anisotropic dispersion dependencies of all four longitudinal waves appearing from equation (\ref{SUSD Obl Longit disp eq general}).

Fig. (\ref{SUSD Obl F5}) allows to compare behavior of two new branches of dispersion dependencies. From Figs. (\ref{SUSD Obl F1})-(\ref{SUSD Obl F5}) we see that frequencies of both new branches increase with the increasing of the wave vector module.

Lower branch (see Figs. (\ref{SUSD Obl F1}), (\ref{SUSD Obl F5})) exists at the parallel propagation of waves, but it does not exist at the perpendicular propagation. Frequency of this wave monotonically decreases to $\omega=0$ with increasing of angle $\theta$ from 0 to $\pi/2$. At $\theta=\pi/2$ structure of equation (\ref{SUSD Obl Longit disp eq general}) changes. It simplifies to an equation having two solutions only, instead of four solutions of equation (\ref{SUSD Obl Longit disp eq general}).

The second (upper) new branch has minimal frequency $\omega_{min}=\mid\Omega\mid$, which is the electron cyclotron frequency (see Figs. (\ref{SUSD Obl F3}) and (\ref{SUSD Obl F5})).

The upper branch shows different behavior than the lower branch. Its frequency also increases with increasing of the wave vector module. However it reaches maximal value at $\theta=\pi/2$. Upper branch frequency decreases with decreasing of angle $\theta$. The upper branch disappears at $\theta=0$. At $\theta=0$ structure of general dispersion equation (\ref{SUSD Obl Longit disp eq general}) changes and we obtain equation (\ref{SUSD Obl Longit disp eq}), which has two solutions only. The upper branch has no trace in equation (\ref{SUSD Obl Longit disp eq}).

In paper \cite{Andreev spin-up and spin-down 1405} it was shown that separated spin evolution and different Fermi pressure for two species of electrons lead to extra term in dispersion dependence of the Langmuir wave propagating parallel and perpendicular to an external magnetic field.

In this paper we study oblique propagation of the Langmuir wave. We also consider properties of the Trivelpiece--Gould wave existing at oblique propagation. We are interested in consideration of described effects in dispersion of these waves. Figs. (\ref{SUSD Obl F6}) and (\ref{SUSD Obl F7}) show contribution of these effects in dispersion surfaces of the Langmuir wave and the Trivelpiece--Gould wave. The lower surface on Fig. (\ref{SUSD Obl F6}) presents usual dispersion dependence of the Langmuir wave $\omega_{0(\uparrow\nearrow)}^{2}=\frac{1}{2}\Biggl[\omega_{Le}^{2}+\Omega^{2}+\frac{1}{3}v_{Fe}^{2}k^{2}
+\sqrt{\biggl(\omega_{Le}^{2}+\Omega^{2}+\frac{1}{3}v_{Fe}^{2}k^{2}\biggr)^{2}
-4\Omega^{2}\biggl(\omega_{Le}^{2}+\frac{1}{3}v_{Fe}^{2}k^{2}\biggr)\cos^{2}\theta}\Biggr]$ $\rightarrow$$\omega_{0\parallel}^{2}=\omega_{Le}^{2}+\frac{1}{3}v_{Fe}^{2}k^{2}$, where $\omega_{0\parallel}$ is the frequency of the Langmuir wave propagating parallel to the external field $\theta=0$, and $\omega_{0(\uparrow\nearrow)}$ is the frequency of the oblique propagating Langmuir wave. The upper surface describes dispersion of Langmuir wave obtained in this paper, which is also presented on Fig. (\ref{SUSD Obl F4}). Upper surface gives dispersion of the Langmuir wave at separated spin evolution and different Fermi pressure for two species of electrons.

The effects under discission give a small contribution in dispersion of the Trivelpiece--Gould wave revealing in increasing of the frequency. The shift of dispersion surface $\Delta\xi=(\omega^{2}_{new}-\omega^{2}_{old})/\Omega^{2}$ is depicted on Fig. (\ref{SUSD Obl F7}). We see that maximal shift appears at $\theta=\pi/4$. This shift increases with increasing of the wave vector module. The shift disappears at $\theta\rightarrow0$ and $\theta\rightarrow\pi/2$.

In this section we have numerically described behavior of the four longitudinal waves existing in magnetized degenerate spin-1/2 plasmas. Some analytical results for limit cases of waves propagating parallel and perpendicular to the external magnetic field can be found in Ref. \cite{Andreev spin-up and spin-down 1405}.

\section{Conclusions}

We have presented the QHD model of spin-1/2 quantum plasmas, where spin-up and spin-down electrons are considered as two different species. This model contains the continuity and Euler equations for each species. Structure of these equations differs from structure of similar equations in QHD with electrons considered as a single species. Particularly we should mention that extra non-linear terms appear in the SSE-QHD equations related to un-conservation of numbers of spin-up and spin-down electrons.

The SSE-QHD also contains equations for evolution of the spin density projections $S_{x}$ and $S_{y}$ on directions perpendicular to the external magnetic field. Projection of the spin density on the direction of the external magnetic field $S_{z}$ is not an independent variable. It appears as difference of concentrations of spin-up and spin-down electrons $S_{z}=n_{\uparrow}-n_{\downarrow}$.

All projections of the spin density $\textbf{S}$ are simultaneously related to both species of electrons. The concentrations $n$ and velocity fields $\textbf{v}$ wear subindexes $\uparrow$ (for spin-up) and $\downarrow$ (for spin-down), but the spin density does not wear them.

The SSE-QHD model arises as a rigorous consequence of the Pauli equation.

Being placed in an external magnetic field a system of degenerate electrons (ions are considered to be motionless, they create positively charged background) has different distributions of spin-up and spin-down electrons. Consequently the Fermi pressure is different for each species.

Account of this effect in the SSE-QHD reveals in existence of two new longitudinal waves in magnetized plasmas.

At consideration of limit cases of wave propagation parallel and perpendicular to the external field we have only one new longitudinal solution existing along with the Langmuir wave. One of two new waves reveals at parallel propagation, and another one exists at perpendicular propagation. Considering oblique propagation we have both new waves existing together with the Langmuir and the Trivelpiece--Gould waves.

We have described described dispersion properties of new waves at oblique propagation. We have studied changes of dispersion of the Langmuir and the Trivelpiece--Gould waves appearing due to different distributions of degenerate spin-up and spin-down electrons in the external magnetic field.

Separated spin evolution QHD shows itself as an useful tool for research of quantum plasmas in magnetic fields. It allows to discower new phenomenon in linear regime of small amplitude perturbations. It also opens possibilities for discovering of new non-linear phenomenon.


\begin{thebibliography}{17}


\bibitem{Maksimov VestnMSU 2000} L. S. Kuz'menkov, S. G. Maksimov, and V. V. Fedoseev,
Vestn. Mosk. Univ., Ser. 3: Fiz., Astron., No. 5, 3 (2000) [Moscow
Univ. Phys. Bull., No. 5, 1 (2000)].

\bibitem{Maksimov Izv 2000} L. S. Kuz'menkov, S. G. Maksimov, and V. V. Fedoseev, Russian Phys. Jour. \textbf{43}, 718 (2000).


\bibitem{MaksimovTMP 2001} L. S. Kuz'menkov, S. G. Maksimov, and V. V. Fedoseev, Theor.
Math. Fiz. \textbf{126} 136 (2001) [Theoretical and Mathematical
Physics, \textbf{126} 110 (2001)].

\bibitem{Marklund PRL07} M. Marklund and G. Brodin,
Phys. Rev. Lett. \textbf{98}, 025001 (2007).

\bibitem{Brodin NJP 07} G. Brodin and M. Marklund, New J. Phys, \textbf{9}, 277
(2007).

\bibitem{Andreev RPJ 07} P. A. Andreev and L. S. Kuz'menkov, Russian Phys. Jour. \textbf{50}, 1251 (2007).

\bibitem{Vagin Izv RAN 06} D. V. Vagin, N. E. Kim, P. A. Polyakov, A. E. Rusakov,
Izvestiya RAN (Proceedings of Russian Academy of science) \textbf{70}, 443 (2006).


\bibitem{Andreev VestnMSU 2007} P. A. Andreev, L.S. Kuz'menkov,
Moscow University Physics Bulletin \textbf{62}, N.5, 271 (2007).

\bibitem{Andreev AtPhys 08} P. A. Andreev, L. S.  Kuz'menkov,
Physics of Atomic Nuclei \textbf{71}, N.10, 1724 (2008).


\bibitem{Misra JPP 10} A. P. Misra, G. Brodin, M. Marklund and P. K. Shukla, J. Plasma Physics \textbf{76}, 857 (2010).

\bibitem{Andreev IJMP 12} P. A. Andreev, L. S. Kuz'menkov,  Int. J. Mod. Phys. B \textbf{26} 1250186 (2012).



\bibitem{Andreev spin-up and spin-down 1405} P. A. Andreev, arXiv:1405.0719.

\bibitem{Andreev 1403 exchange} P. A. Andreev, arXiv:1403.6075.
\bibitem{Trukhanova 1405 exchange} Mariya Iv. Trukhanova and Pavel A. Andreev, arXiv:1405.6294.


\bibitem{Iqbal JPP 13} Mubashar Iqbal, J. Plasma Physics, \textbf{79}, 19 (2013).
\bibitem{Shahid PP 12} M. Shahid, D. B. Melrose, M. Jamil, and G. Murtaza, Phys. Plasmas, \textbf{19}, 112114 (2012).







\bibitem{Shukla PhUsp 2010} P. K. Shukla, B. Eliasson, Phys. Usp. \textbf{53}, 51 (2010).

\bibitem{Shukla RMP 11} P. K. Shukla, B. Eliasson, Rev. Mod. Phys. \textbf{83}, 885 (2011).




\bibitem{Uzdensky arxiv review 14} D. A. Uzdensky and S. Rightley,
Reports on Progress in Physics, \textbf{77}, Issue 3, 036902 (2014). 


\bibitem{Andreev Asenjo 13} P. A. Andreev,
F. A. Asenjo, and S. M. Mahajan, arXiv: 1304.5780.


\bibitem{Marklund PLA 11} M. Marklund, P. J. Morrison, Physics Letters A \textbf{375}, 2362 (2011).

\bibitem{Mahajan PL A 13} S. M. Mahajan, F. A. Asenjo, Phys. Lett. A \textbf{377}, 1430 (2013).
\bibitem{Trukhanova PrETP 13} M. I. Trukhanova,  Prog. Theor. Exp. Phys., 111I01 (2013).

\bibitem{Asenjo PL A 12} F. A. Asenjo, Phys. Lett. A \textbf{376}, 2496 (2012).

\bibitem{Bychkov PP 10} Vitaly Bychkov, Mikhail Modestov, and Mattias Marklund, Phys. Plasmas \textbf{17}, 112107 (2010).


\bibitem{Mushtaq PP 10} A. Mushtaq, and S. V. Vladimirov, Phys. Plasmas \textbf{17}, 102310 (2010).
\bibitem{Mushtaq PP EPJD 11} A. Mushtaq, and S. V. Vladimirov, Eur. Phys. J. D \textbf{64}, 419 (2011).



\bibitem{Mushtaq PP 12} A. Mushtaq, R. Maroof, Zulfiaqr Ahmad, and A. Qamar, Phys. Plasmas \textbf{19}, 052101
(2012).

\bibitem{Chang Li PP 14} Sheng-Chang Li and Jiu-Ning Han, Phys. Plasmas \textbf{21}, 032105 (2014).









\bibitem{Shahid PP 13} M. Shahid and G. Murtaza,
Phys. Plasmas \textbf{20}, 082124 (2013).

\bibitem{Sharma PP 14} Prerana Sharma and R. K. Chhajlani, Phys. Plasmas \textbf{21}, 032101 (2014).


\bibitem{Hussain PLA 13} A. Hussain, Z. Iqbal, G. Brodin, G. Murtaza, Phys. Lett. A \textbf{377}, 2131 (2013).









\bibitem{MaksimovTMP 1999} L. S. Kuz'menkov and S. G. Maksimov,  Teor. i Mat. Fiz.,
\textbf{118} 287 (1999) [Theoretical and Mathematical Physics \textbf{118} 227 (1999)].


\bibitem{Andreev PRB 11} P. A. Andreev, L. S. Kuzmenkov, M. I. Trukhanova, Phys. Rev. B \textbf{84}, 245401 (2011).

\bibitem{Takabayasi PTP 55 a} T. Takabayasi, Prog. Theor. Phys. \textbf{13}, 222
(1955).

\bibitem{Takabayasi PTP 54} T. Takabayasi, Prog.
Theor. Phys. \textbf{12}, 810 (1954).

\bibitem{Takabayasi PTP 55 b} T. Takabayasi, Prog. Theor. Phys. \textbf{14}, 283 (1955).



%
\end{thebibliography}
\end{document}